\documentclass[preprint,preprintnumbers,amsmath,amssymb]{revtex4}

\usepackage{graphicx}
\usepackage{dcolumn}
\usepackage{bm}
\usepackage{epsfig}

\begin{document}


\title{Critical surfaces for general bond percolation problems}
\author{Christian R. Scullard}
\email{scullard@uchicago.edu}
\affiliation{Department of Geophysical Sciences, University of Chicago, Chicago, Illinois 60637, USA}
\author{Robert M. Ziff}
\email{rziff@umich.edu}
\affiliation{Michigan Center for Theoretical Physics and Department of Chemical Engineering, University of Michigan, Ann Arbor, Michigan 48109-2136, USA}
\date{today}

\begin{abstract}
We present a general method for predicting bond percolation thresholds
and critical surfaces for a broad class of two-dimensional periodic
lattices, reproducing many known exact results and providing
excellent approximations for several unsolved lattices. For the checkerboard and inhomogeneous bow-tie lattices, the method yields predictions that agree with numerical measurements to more
than six figures, and are possibly exact.
\end{abstract}

\maketitle
Percolation deals with the formation of long-range connectivity in a
random network, represented by a given graph. In bond percolation, we define every edge of a graph to be open with probability $p$ and closed with probability $1-p$. The average size of resulting clusters of open bonds becomes greater with increasing $p$ and at a special value, $p_c$, called the critical threshold, an infinite cluster appears. The value of $p_c$ strongly depends on the lattice under consideration.

\begin{center}
\begin{figure}
 \includegraphics{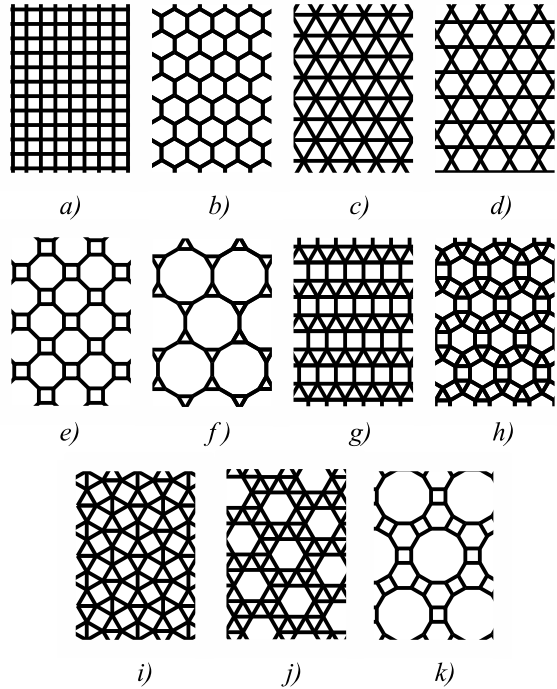} 
\caption{The Archimedean lattices; a) square $(4^4)$; b) honeycomb $(6^3)$; c) triangular $(3^6)$; d) kagome $(6,3,6,3)$; e) $(4,8^2)$; f) $(3,12^2)$; g) $(3^3,4^2)$; h) $(3,4,6,4)$; i) $(3^2,4,3,4)$; j) $(3^4,6)$; k) $(4,6,12)$ .} \label{fig:archimedean}
\end{figure}
\end{center}

Knowledge of the percolation thresholds is central
to understanding and applying this process, and the problem
of deriving or computing those thresholds has been the subject
of intense study since the introduction of the model over 50
years ago \cite{BroadbentHammersley,Flory}. Even after so many years of research there still exists no general method for deriving critical probabilities of arbitrary graphs, although schemes of varying accuracy have been proposed \cite{Galam,Wierman2005,Neher}. All non-trivial exact solutions are confined to two dimensions and special lattice classes \cite{Ziff06,ChayesLei}; these exact thresholds all appear as the roots of polynomials with integer coefficients. Although we focus mainly on the Archimedean lattices in this Letter (Fig.\ \ref{fig:archimedean}), we present a method whereby a polynomial can, in principle, be associated with any two-dimensional periodic lattice and the prediction for the bond threshold found as the root in $[0,1]$. Our results agree with all exact solutions, and for unsolved problems we find excellent approximations for every system considered. In two cases, the checkerboard and inhomogeneous bow-tie
lattices, we find predictions that agree with our numerical tests
to high precision and these formulas may be exact.
\begin{center}
\begin{figure}
\includegraphics{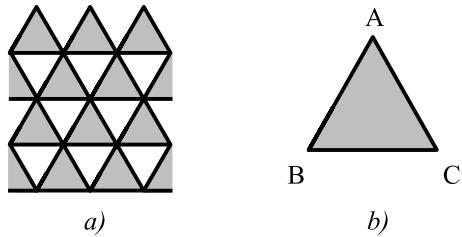} 
\caption{a) A class of exactly solvable lattices; b) every shaded triangle of a) can represent any network of sites and bonds contained in the vertices $(A,B,C)$.} \label{fig:selfdual}
\end{figure}
\end{center}

\begin{center}
\begin{figure}
 \includegraphics{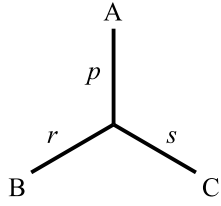} 
\caption{Using this star in Figure \ref{fig:selfdual}b results in the honeycomb lattice (Fig. \ref{fig:archimedean}b).} \label{fig:star}
\end{figure}
\end{center}

\begin{center}
\begin{figure}
 \includegraphics{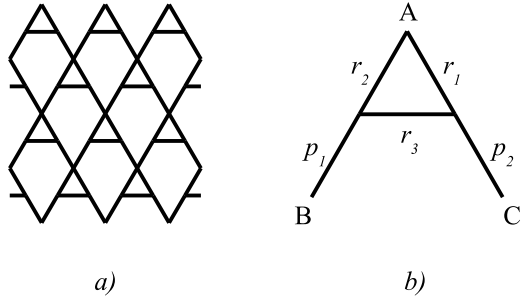} 
\caption{a) The martini-A lattice; b) The assignment of probabilities on the unit cell.} \label{fig:Alattice}
\end{figure}
\end{center}

All known exact thresholds follow from duality arguments.
Using a general triangle-triangle transformation \cite{Ziff06}, which is an extension of
the well-known star-triangle transformation \cite{SykesEssam}, broad classes of exact thresholds and
critical surfaces have been derived recently \cite{ZiffScullard06}. The approach works only for
lattices in which the basic repeated cell can be contained in a triangle, and the
triangles are arranged in a self-dual way, as shown for example in Fig. \ref{fig:selfdual}a. Referring to the generalized cell in Figure\ \ref{fig:selfdual}b, the critical threshold is found by using the condition \cite{Ziff06, ChayesLei}:
\begin{equation}
P(A,B,C)=P(\bar{A},\bar{B},\bar{C}) \label{eq:critcond}
\end{equation}
where $P(A,B,C)$ is the probability that the vertices $(A,B,C)$ are connected by open bonds and $P(\bar{A},\bar{B},\bar{C})$ is the probability that none are connected. As a simple example, we take for Figure \ref{fig:selfdual}b the star with inhomogeneous probabilities assigned as in Figure\ \ref{fig:star}, which results in the honeycomb lattice (Figure\ \ref{fig:archimedean}b). By allowing distinct probabilities for the bonds on the unit cell, we find a critical surface as opposed to a critical point, and application of (\ref{eq:critcond}) gives \cite{SykesEssam}
\begin{equation}
H(p,r,s) \equiv prs - rp - rs - ps + 1=0 .\label{eq:hex}
\end{equation}
We locate the critical probability for the homogeneous system by setting $r=s=p$:
\begin{equation}
 p^3-3 p^2+1=0
\end{equation}
so $p_c=1-2 \sin \pi/18 \approx 0.652704$ \cite{SykesEssam}. If we want to find the threshold of the ``martini-A'' lattice (Fig.\ \ref{fig:Alattice}a), we use the A cell shown in Figure \ref{fig:Alattice}b, which leads to
\begin{eqnarray}
A(p_1,p_2,r_1,r_2,r_3) &\equiv& 1 -  p_1 r_2 - p_2 r_1 - p_1 p_2 r_3   \nonumber \\
&-& p_1 r_1 r_3 - p_2 r_2 r_3 + p_1 p_2 r_1 r_3 \nonumber \\
&+& p_1 p_2 r_2 r_3 + p_1 r_1 r_2 r_3 \nonumber \\
&+& p_2 r_1 r_2 r_3 - p_1 p_2 r_1 r_2 r_3 =0 . 
\end{eqnarray}
Setting all probabilities equal leads to the homogeneous polynomial and the bond threshold $p_c=0.625457...$ \cite{Ziff06}. Problems that can be solved with equation (\ref{eq:critcond}) have critical surface formulas of the same basic form as these two examples. For a lattice of the type in Figure \ref{fig:selfdual} in which each of the $n$ bonds of the unit cell is assigned a different probability $p_i$, this form is
\begin{equation}
 f(p_1,p_2,...,p_n)=0 \nonumber
\end{equation}
where the function $f$ is at most first-order in each argument. We refer to this latter property of $f$ as ``linearity''. Since equation (\ref{eq:critcond}) compares events that happen only on a unit cell, it is clear that any $f$ found through (\ref{eq:critcond}) will have this property. A consequence of linearity is that the critical threshold for a homogeneous system is the root of a polynomial of order at most $n$.

\begin{center}
\begin{figure}
 \includegraphics{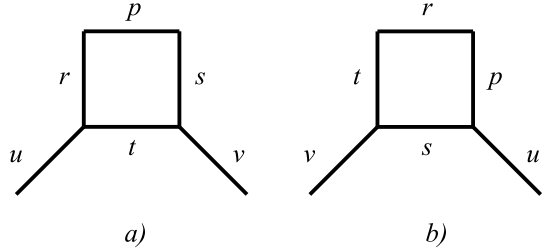} 
\caption{a) The assignment of probabilities on the $(4,8^2)$ unit cell; b) A ``re-partitioning'' of the graph into different unit cells.} \label{fig:foureightsquared}
\end{figure}
\end{center}

To construct the critical surface for a general periodic lattice, our basic assumption is that linearity holds in general, even for lattices that cannot be solved with (\ref{eq:critcond}). In order to find $f$ for a given lattice, we simply write the most general function that is first order in all its arguments, and then impose symmetries and agreement with special cases until the unknown coefficients are constrained. As an illustrative example, we choose the $(4,8^2)$ lattice of Figure \ref{fig:archimedean}e with the probabilities assigned on the unit cell as in Figure \ref{fig:foureightsquared}a. The critical function can be written
\begin{equation}
 F(p,r,s,t,u,v)\equiv \sum_{i=0}^1 ... \sum_{n=0}^1 a_{ijklmn} p^i r^j s^k t^l u^m v^n \label{eq:generalf}
\end{equation}
where there are $64$ $a$'s to be determined. In general, with a unit cell of $n$ bonds there are $2^n$ coefficients. Since (\ref{eq:generalf}) can be multiplied by an arbitrary factor without changing the threshold, we are free to choose one of the $a$'s, and we set the constant term to $+1$ as a matter of convention. To determine the rest, we impose symmetries and consider special cases. Since there are so many coefficients, we must use a computer algebra program to impose the various constraints. We start with the repartitioning symmetry shown in Figure \ref{fig:foureightsquared}b. That is, there is more than one way to divide the $(4,8^2)$ lattice into unit cells and different choices should not alter the critical function, so
\begin{equation}
 F(p,r,s,t,u,v)=F(r,t,p,s,v,u) . \label{eq:repart}
\end{equation}
This constraint requires many of the $a$'s to be equal and reduces the number to be determined to $21$. Next, we impose mirror symmetry; reflecting the lattice about a vertical line through the center of the unit cell does not change the threshold. This means
\begin{equation}
 F(p,r,s,t,u,v)=F(p,s,r,t,v,u) \nonumber
\end{equation}
which reduces the number of coefficients by $3$, to $18$. Using a different repartitioning besides (\ref{eq:repart}) and reflecting about a horizontal line do not give any additional constraints. We now turn to special cases. If we remove the $t$ bond by setting its probability to $0$ we are left with a honeycomb lattice with two doubled bonds. Therefore, we require
\begin{equation}
 F(p,r,s,0,u,v)=H(p,ru,sv) \nonumber
\end{equation}
which leaves only three undetermined $a$'s. If we set $p=1$ we get the martini-A lattice, so
\begin{equation}
 F(1,r,s,t,u,v)=A(s,r,t,u,v) \nonumber
\end{equation}
which fixes the remaining coefficients. We finally arrive at the critical surface,
\begin{eqnarray}
F(p,r,s,t,u,v)&=&1 - p r u - s t u - p s v - r t v \nonumber \\ 
&-& r s u v - p t u v + p r s u v + p t r u v  \nonumber \\
&+& p s t u v + r s t u v + p r s t u \nonumber \\
&+& p r s t v - 2 p r s t u v=0 .\label{eq:foureightsquared}
\end{eqnarray}
The polynomial for the homogeneous case is
\begin{equation}
F(p,p,p,p,p,p)= 1 - 4 p^3 - 2 p^4 + 6 p^5 - 2 p^6=0 \label{eq:fespoly}
\end{equation}
with the solution $p_c=0.676835...$ on $[0,1]$. Parviainen's numerical estimate for this threshold is $p_c=0.67680232(63)$, where the number in parentheses is the standard error in the last digit(s). Although our solution is ruled out, it is accurate to four significant figures. Additionally, we have an analytic expression for the full critical surface, something that would be practically impossible to obtain numerically. While there are alternative paths to derive (\ref{eq:foureightsquared}) that
are more efficient, the steps presented here demonstrate the general approach used for more complicated lattices. In fact, we have employed it to obtain critical surfaces and polynomials for all but two of the Archimedean lattices. These results are shown in Table \ref{table:bondthresholds} along with the rigorous confidence intervals of Riordan and Walters \cite{Riordan} and the numerical values of Parviainen \cite{Parviainen}. Note that our estimate for the kagome lattice matches the 1979 conjecture of Wu \cite{Wu79}, who took a similar approach in an attempt to solve that problem.

Although all of our thresholds for the Archimedean lattices agree to at least four significant figures with Parviainen's results, they are ruled out numerically, implying that the linearity hypothesis does not hold exactly in these cases.
\begin{table}
\begin{center}
\begin{tabular}{clclclc}
lattice & & $p_c^{\mathrm{approx}}$ & & $p_c^{\mathrm{num}}$& & bounds$^c$\\
\hline
kagome &\vline& $0.5244297^d$&\vline&$0.5244053(3)^a$&\vline&$[0.52415,0.52465]$\\
$(3,12^2)$ &\vline& $0.7404233^e$ &\vline& $0.7404220(8)^b$&\vline&$[0.7402,0.7407]$\\
$(4,8^2)$ &\vline& $0.676835^f$&\vline&$0.6768023(6)^b$&\vline&$[0.6766,0.6770]$\\
$(3^3,4^2)$ &\vline& $0.419615^g$&\vline&$0.4196419(4)^b$&\vline&$[0.4194,0.4199]$\\
$(3,4,6,4)$ &\vline&$0.524821^h$&\vline&$0.5248326(5)^b$&\vline&$[0.5246,0.5251]$\\
$(3^2,4,3,4)$ &\vline&$0.414120^i$ &\vline& $0.4141374(5)^b$&\vline&$[0.4139,0.4144]$\\
\end{tabular}
\end{center}
\caption{Comparison of bond percolation estimates with numerical results and the Riordan and Walters confidence intervals; a: \cite{ZiffSuding97}, b: \cite{Parviainen}, c: \cite{Riordan}, d: $1 - 3 p^2 - 6 p^3 + 12 p^4 - 6 p^5 + p^6=0$, e: $1 - 3 p^4 - 6 p^5 + 3 p^6 + 15 p^7 - 15 p^8 + 4 p^9=0$, f: Eq.\ (\ref{eq:fespoly}), g: $1 - 4 p^2 - 12 p^3 + 104 p^5 
- 193 p^6 + 146 p^7 - 45 p^8 + 2 p^{10}=0$, h: $1 - 6 p^3 - 12 p^4 - 6 p^5 + 69 p^6 + 60 p^7 - 363 p^8 + 448 p^9 - 252 p^{10} + 66 p^{11} - 6 p^{12} =0$, i: $1 - 4 p^2 - 12 p^3 - 2 p^4 + 106 p^5 - 186 p^6 + 132 p^7 - 36 p^8 - 2 p^9 + 2 p^{10}=0$ .} \label{table:bondthresholds}
\end{table}
However, it is possible to get successively better estimates with this method by increasing the number of distinct probabilities. Although there is a minimum number that one can choose, corresponding to the size of the unit cell, if we extend the inhomogeneity to a neighboring cell we get a refinement of the approximation. For the $(3^3,4^2)$ lattice for example, the simplest unit cell contains five bonds, and using five probabilities leads to a polynomial ($1 - 2 p - 2 p^2 + 3 p^3 - p^4$) with root in $[0,1]$, $p_c=0.419308...$ . However, the value in Table \ref{table:bondthresholds} is based on covering two unit cells by using ten probabilities, giving a tenth-order polynomial. This estimate agrees much better with the numerical result and falls within the confidence interval of Ref. \cite{Riordan}. In general, however, it becomes increasingly difficult to constrain systems with greater numbers of distinct probabilities, since the set of coefficients quickly becomes large. As such, all other results in Table \ref{table:bondthresholds} are based on a single unit cell. In fact, we have yet to find expressions for even the single-cell critical surfaces of the $(3^4,6)$ and $(4,6,12)$ lattices because of the large numbers of bonds involved.

\begin{center}
\begin{figure}
 \includegraphics{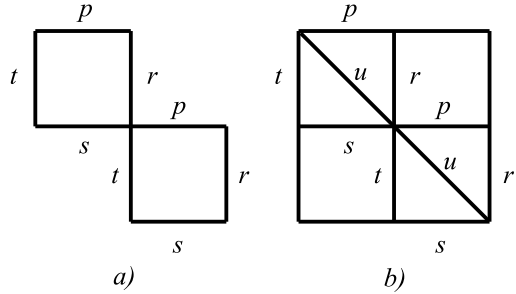} 
\caption{a) The inhomogeneous square lattice with the checkerboard assignment of probabilities; b) the assignment of probabilities on the 5-bond inhomogeneous bow-tie lattice (see Figure \ref{fig:duality}).} \label{fig:checker-bowtie}
\end{figure}
\end{center}

Our method also gives some solutions that agree with simulations to very high precision, and may be exact. If we set $u=v=1$ in (\ref{eq:foureightsquared}) we get the square lattice (see Fig. \ref{fig:checker-bowtie}a) but with four different probabilities. Equation (\ref{eq:foureightsquared}) predicts
\begin{eqnarray}
C(p,r,s,t) &\equiv& 1 - p r - p s - r s - p t - r t - s t \nonumber \\
&+& p r s + p r t + p s t + r s t =0 . \label{eq:checkerboard}
\end{eqnarray}
Wu \cite{Wu82} considered this situation, which he called the ``checkerboard'', in the context of the $q-$state Potts model, for which percolation is the limit as $q \rightarrow 1$. Equation (\ref{eq:checkerboard}) is equivalent to his formula in that limit. Although Wu's conjecture was shown to be incorrect for $q=3$ \cite{Enting},
it holds in the $q=2$ (Ising) case, and its validity for $q = 1$ (percolation) evidently has
not been tested.  We
have carried out high-precision numerical simulations, using the gradient percolation method described in \cite{ZiffSapoval}, which
are consistent with (\ref{eq:checkerboard}) being exact. To test predictions of the critical surface, we fix some probabilities and find the critical value for a single remaining parameter. For example, if we set $p=73/90$, and assume $t=s=r$, then (\ref{eq:checkerboard}) predicts $r_c=0.4$. Numerically, we find $r_c = 0.40000004(10)$ simulating a total of $10^{14}$ frontier bonds in lattices of various gradients. We also found similar agreement at other parameter values. Note that (\ref{eq:checkerboard}) cannot be derived from condition (\ref{eq:critcond}).

\begin{center}
\begin{figure}
 \includegraphics{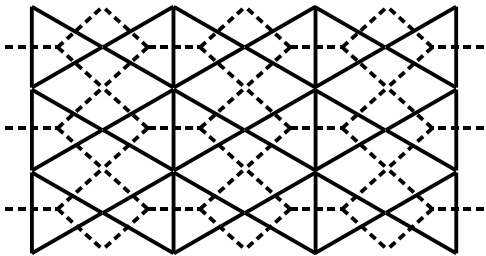} 
\caption{The duality transformation for the bow-tie
lattice (solid lines). The dotted graph is the dual.} \label{fig:duality}
\end{figure}
\end{center}
If we set $v=1$ in (\ref{eq:foureightsquared}), we get a prediction for the
dual of the bow-tie lattice \cite{Wierman84} (Figure \ref{fig:duality}). The bond thresholds of dual pairs of lattices are related \cite{SykesEssam} by
$p_c(L)=1-p_c(L_d)$. This can be easily generalized for critical surfaces. Specifically, we find the $5$-bond version (Figure \ref{fig:checker-bowtie}b) of the bow-tie critical surface as
\begin{equation}
F(1-p,1-r,1-s,1-t,1-u,1)=0 . \nonumber
\end{equation}
We thus find, for the lattice of Figure \ref{fig:checker-bowtie}b,
\begin{eqnarray}
B(p,r,s,t,u)&\equiv&1 - u - p r - p s - r s - p t \nonumber \\
 &-& r t - s t + p r s + p r t  + p s t \nonumber \\
 &+& r s t+ p r u + s t u - p r s t u = 0 \label{eq:bowtie}
\end{eqnarray}
which reduces to (\ref{eq:checkerboard}) in the limit $u=0$. Setting the probabilities equal yields
\begin{equation}
B(p,p,p,p,p)= 1 - p - 6 p^2 + 6 p^3 - p^5=0 \nonumber
\end{equation}
with solution $p_c=0.404518...$, in agreement with Wierman's exact result, \cite{Wierman84}. We studied the situation in Figure \ref{fig:checker-bowtie}b numerically and once again find agreement to high precision. For example, if we set $u=r=s=p$ and $t=1/2$, then (\ref{eq:bowtie}) gives $p_c=(3 - \sqrt{5})/2 = 0.38196601$. Numerically, we find $p_c=0.3819654(5)$, which is well within two standard deviations of the prediction. Like (\ref{eq:checkerboard}), (\ref{eq:bowtie}) cannot be derived using (\ref{eq:critcond}).

Thus, in conclusion, we have a method that gives very
 accurate approximations to the bond critical surfaces for many
lattices, and in two cases -- the checkerboard
and inhomogeneous bow-tie lattices -- formulas that are consistent with precise numerical results. Although the procedure presented here would work in principle for site percolation, there are not many exact site critical thresholds to which the Archimedean lattices can reduce in special cases. This problem also limits the extension to higher dimensions, where there are no exact solutions of any kind.

RMZ acknowledges support of the National Science Foundation under grant DMS-0553487.

\bibliography{scullard_ziffv5}

\end{document}